\documentclass[aip,apl,reprint]{revtex4-1}

\usepackage{physics}
\usepackage{graphicx}
\usepackage{color}
\usepackage[colorinlistoftodos]{todonotes}

\newcommand{\imwidth}{0.91\columnwidth}

\begin{document}

\title{Active tuning of high-Q dielectric metasurfaces}

\author{Matthew Parry}
\email[Address correspondence to: ]{Matthew.Parry@anu.edu.au}
\author{Andrei Komar}
\author{Ben Hopkins}
\affiliation{Nonlinear Physics Centre and Centre for Ultrahigh Bandwidth Devices for Optical Systems (CUDOS), Research School of Physics and Engineering, Australian National University, Canberra, ACT 2601, Australia}

\author{Salvatore Campione}
\author{Sheng Liu}
\affiliation{Center for Integrated Nanotechnologies, Sandia National Laboratories, Albuquerque, New Mexico 87185, United States}

\author{Andrey E. Miroshnichenko}
\affiliation{Nonlinear Physics Centre and Centre for Ultrahigh Bandwidth Devices for Optical Systems (CUDOS), Research School of Physics and Engineering, Australian National University, Canberra, ACT 2601, Australia}

\author{John Nogan}
\author{Michael B. Sinclair}
\author{Igal Brener}
\affiliation{Center for Integrated Nanotechnologies, Sandia National Laboratories, Albuquerque, New Mexico 87185, United States}

\author{Dragomir N. Neshev}
\affiliation{Nonlinear Physics Centre and Centre for Ultrahigh Bandwidth Devices for Optical Systems (CUDOS), Research School of Physics and Engineering, Australian National University, Canberra, ACT 2601, Australia}

\begin{abstract}
We demonstrate the active tuning of all-dielectric metasurfaces exhibiting high-quality factor (high-Q) resonances. The active control is provided by embedding the asymmetric silicon meta-atoms with liquid crystals, which allows the relative index of refraction to be controlled through heating. It is found that high quality factor resonances ($Q=270\pm30$) can be tuned over more than three resonance widths.
Our results demonstrate the feasibility of using all-dielectric metasurfaces to construct tunable narrow-band filters.
\end{abstract}

\maketitle


There is rapidly growing interest in all-dielectric metasurfaces~\cite{Kuznetsov16:Sci} for optical applications.  Examples include flat lenses~\cite{Arbabi:2015:NatComm, Khorasaninejad:2016:Sci}, beam converters~\cite{Lin:2014:Sci, Yu:2015:LPOR, Chong2015} and
holograms~\cite{Arbabi:2015:NatNano, Wang:16:Optica,Chong2016}. Dielectric metasurfaces are two-dimensional arrangements of dielectric 
nano-resonators which exhibit low absorption in the visible and infra-red spectral range. These low losses allow one to obtain resonances with a high quality-factor (high-Q) in comparison with their plasmonic counterparts. Importantly, the interference of electric and magnetic Mie-type resonant modes of the same strength allows for fundamentally new effects to be realized, such as unidirectional scattering~\cite{Fu:2013:NatComm} and unity transmission in the Huygens regime~\cite{Decker:2015:ADOM, staude13,Campione2013}. Furthermore, the Fano interference of bright and dark modes in the dielectric resonators~\cite{Wu:2014:NatComm, Yang:2014:NatComm, Jain:2015:ADOM} provides a pathway to obtaining narrow-band resonances with quality-factors of up to several hundred~\cite{campione16}. Such quality factors open up new applications for dielectric metasurfaces, such as high-sensitivity sensors~\cite{Wu:2014:NatComm, Yang:2014:NatComm} and narrow-band filters~\cite{Zhao:15:OE}.

To date, all narrow-band dielectric metasurfaces have been based on static designs, defined through the choice of geometry. Such static design does not offer the flexibility required for many scientific and industrial applications, including tunable narrow band filters. Therefore implementing {\sl dynamic control} over the response of the metasurface is essential. There have been various schemes proposed for tuning dielectric metasurfaces which focus on the tuning of the refractive index, either of the dielectric resonators or of the surrounding environment. These include the electrical~\cite{Iyer:2016:ADOM} or thermal~\cite{Lewi:2017:arXiv} tuning of semiconductor metasurfaces, as well as tuning of the embedding medium~\cite{Wang:15:OME}, using for example liquid crystals (LCs)~\cite{sautter15, Komar:2017:APL}. To date however, the tuning has only been demonstrated with relatively broad resonances. That is, only a low figure of merit (overall spectral shift divided by the resonance width is less than 1) for the tunability has been achieved, so the spectral shifts {\em do not} generally produce a dramatic change in transmission over the tuning range. 

\begin{figure}
    \includegraphics[width=\imwidth]{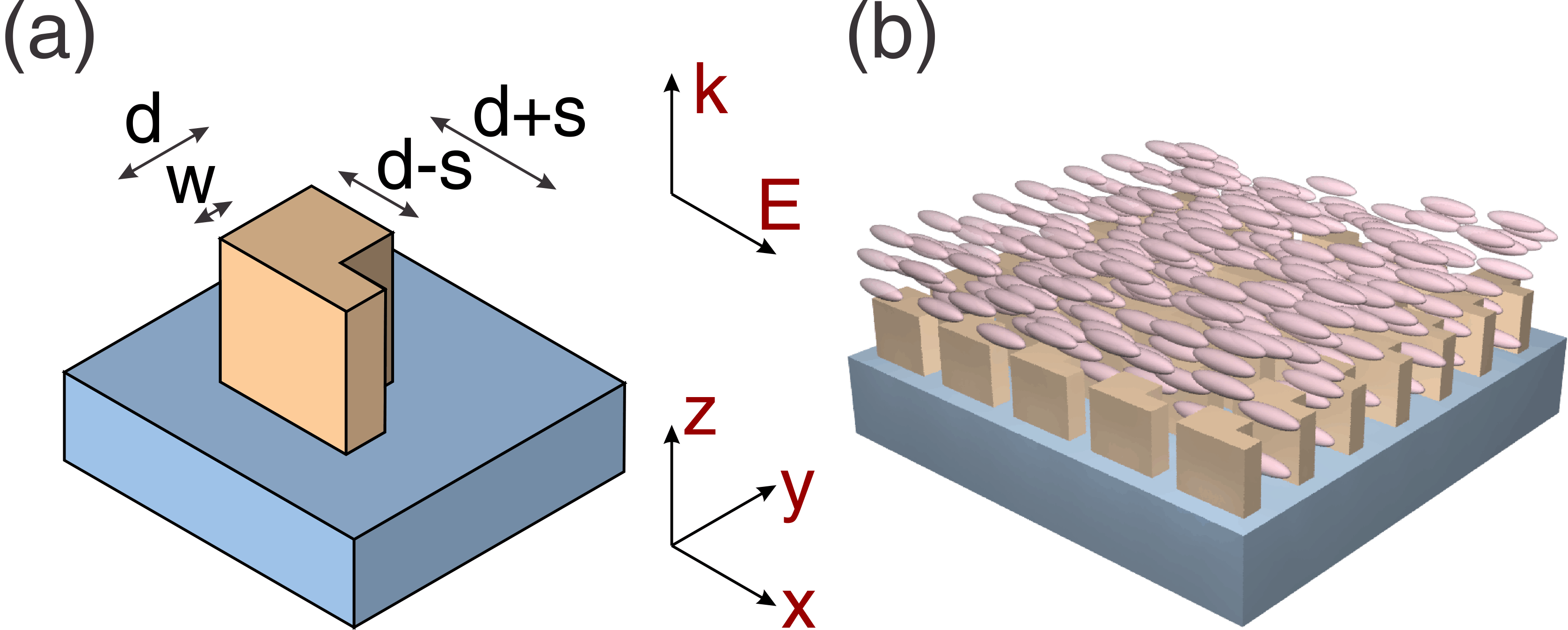}
	\caption{\label{fig:resonator-design}(a) Sketch of the poly-Si resonator on a quartz substrate. The dimensions are $d=300$\,nm, $s=80$\,nm, $w=150$\,nm, periodicity $p=600$\,nm and height $h=250$\,nm. (b) An illustration of the metasurface after infiltration with E7 liquid crystals.} 
\end{figure}

Here, we demonstrate the control of a high-Q ($Q=270\pm30$) dielectric metasurface by changing the properties of the medium surrounding the meta-atoms. 
We utilize nano-resonators with an asymmetric geometry, which allows for weak coupling into their low-radiative multipoles~\cite{campione16} and therefore produce high-Q transmission features. To enable the tuning of these high-Q resonances we embeded the metasurface with nematic LCs, where the LC molecules are aligned along one of the principal axes of the resonators (Fig.~\ref{fig:resonator-design}). By heating the LCs above the isotropic transition temperature, we can tune the narrow band transmission features of the dielectric metasurface by more than three resonance widths, achieving a figure of merit ($\flatfrac{\Delta\lambda}{FWHM}$) of $FOM=3.3\pm0.6$. Tunable dielectric metasurfaces such as these therefore show good promise as tunable narrow band filters.

The all-dielectric frequency selective metasurface used in our experiment consists of a square lattice of asymmetric silicon resonators on a quartz substrate, as shown in Fig.~\ref{fig:resonator-design}(a). When the metasurface is infiltrated with LCs (Fig.~\ref{fig:resonator-design}(b)) the resonance condition of the meta-atoms becomes dependent on the refractive index of the LC medium. Importantly, the LC refractive index can be easily adjusted through either heating or the application of an electric field due to the LCs temperature and electric field dependent anisotropic refractive index~\cite{khoo95}.

\begin{figure}[tb]
  \includegraphics[width=\imwidth]{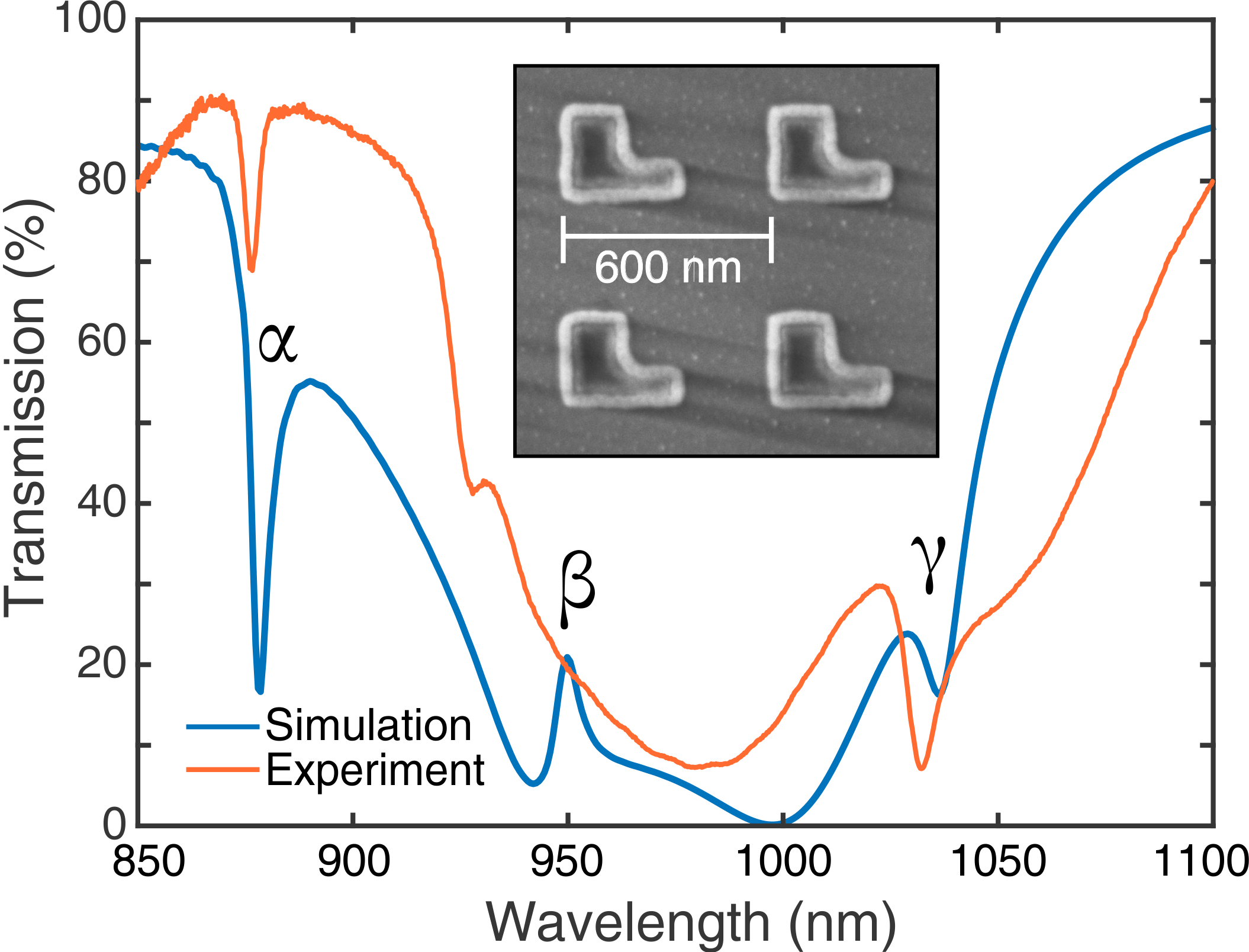}
	\caption{Comparison of the experimentally measured (red curve) and the numerically calculated (blue curve) transmission spectra through the silicon metasurface before the application of the liquid crystals (i.e. superstrate of $n=1$). $\alpha$, $\beta$ \& $\gamma$ mark the three featured resonances discussed below. The inset shows a scanning electron micrograph of the sample.}
    \label{fig:simulation}
\end{figure}

The metasurface was fabricated on a 250\,nm poly-crystalline silicon layer which was deposited using low pressure chemical vapor deposition. The layer was subsequently patterned using electron-beam lithography and reactive-ion etching. A scanning electron micrograph (SEM) image of a typical fabricated surface is shown in the inset of Fig.~\ref{fig:simulation}. The transmission through the fabricated metasurface was measured using a transmission setup~\cite{sautter15,Komar:2017:APL} that was constructed in-house. Simulations of the metasurface's response were done using the finite difference time domain method in Lumerical.

The asymmetry in the resonators was introduced so that \mbox{$\mathrm{E}_x$-polarized} light is able to couple into the longitudinal, $m_z$, magnetic dipole which is parallel to the propagation direction. The $m_z$ dipole radiates dominantly into the plane of the metasurface and so for an infinite array of resonators the energy stored into the excitation of the longitudinal magnetic dipoles is damped only by indirect coupling with the transverse, $p_x$, electric dipole moments of the metasurface~\cite{campione16}. This excitation results in a narrow transmission peak, marked as $\gamma$ in Fig.~\ref{fig:simulation}.

Another two high Q-factor resonances can also be observed in both the measured and calculated transmission spectra of Fig.~\ref{fig:simulation}. The resonance at shorter wavelengths (marked $\alpha$) has a Q-factor of $270\pm30$. Our simulations also reveal a third resonance, labeled $\beta$ in Fig.~\ref{fig:simulation}(blue curve), however this resonance is not well-pronounced in the experimental measurements (red curve), possibly due to fabrication imperfections.

\begin{figure}
   \includegraphics[width=0.99\columnwidth]{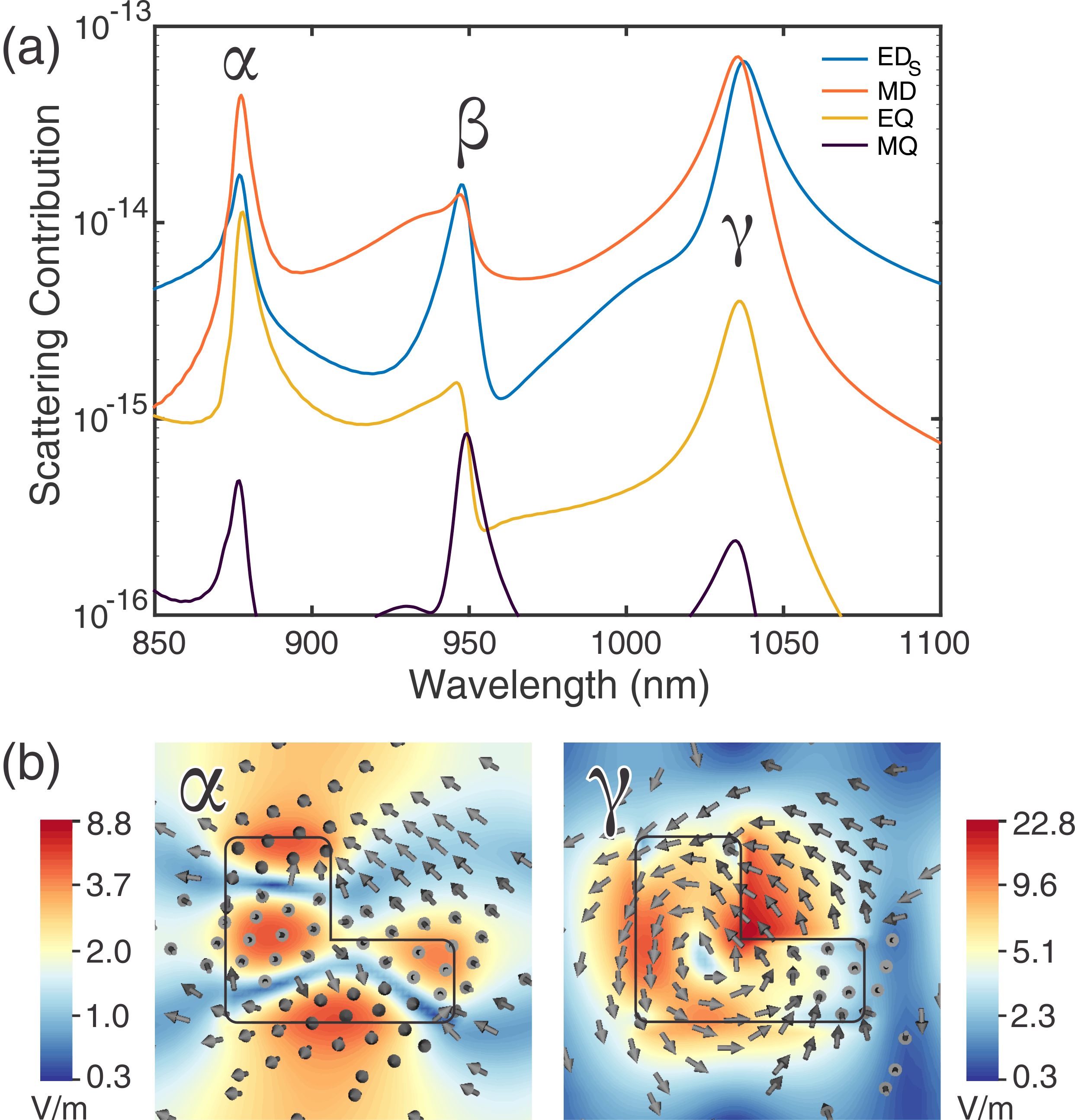}
	\caption{(a) Decomposition of the spectrum in Fig.~\ref{fig:simulation} (Before the application of the liquid crystals) into Spherical Electric Dipole ($ED_S$), Magnetic Dipole (MD), Electric Quadrupole (EQ) and Magnetic Quadrupole (MQ). Note that the scattering contribution is a log scale. (b) The electric field profiles of the $\alpha$ and $\gamma$ resonances.}
     \label{fig:multipoles}
\end{figure}

In order to identify the nature of the observed resonant features, in Fig.~\ref{fig:multipoles} we show the simulated multipolar decomposition of their scattering  into multipoles, assuming air as an isotropic superstrate medium ($n=1$). As can be seen, the $\gamma$ resonance is formed by coupling into the $m_z$ magnetic dipole, as shown by the circulating electric field in Fig.~\ref{fig:multipoles}(b). It has been shown previously that by using illumination which is annular in $k$-space this Fano resonance provides a transmission feature with a quality factor~$\sim 600$~\cite{campione16}. The $\alpha$ resonance, on the other hand, possesses a mixture of several multipolar contributions, with the magnetic dipole dominating, which is caused by the broken symmetry of the meta-atoms. The sharp drop in transmission makes this resonance a candidate for a band-stop filter. We can also see that the $\beta$ resonance, like $\gamma$, involves simultaneous electric and magnetic dipole contributions.
 
Next, we fabricated the LC cell using the metasurface as the base substrate and covered it with a layer of positive dielectric anisotropic LC molecules (E7 LC mixture). The thickness of the cell was set by $5\,\mu$m spherical plastic spacer particles blown onto the metasurface prior to infiltration with E7. To achieve initial orientation of LCs an alignment layer of nylon6 was spin-coated on the upper substrate and rubbed in the $x$-axis direction (Fig.~\ref{fig:resonator-design}(b)). The light was incident from the back of the cell to ensure that the polarization was not affected by the birefringence of the LCs in their nematic state.

\begin{figure}
  \includegraphics[width=0.99\columnwidth]{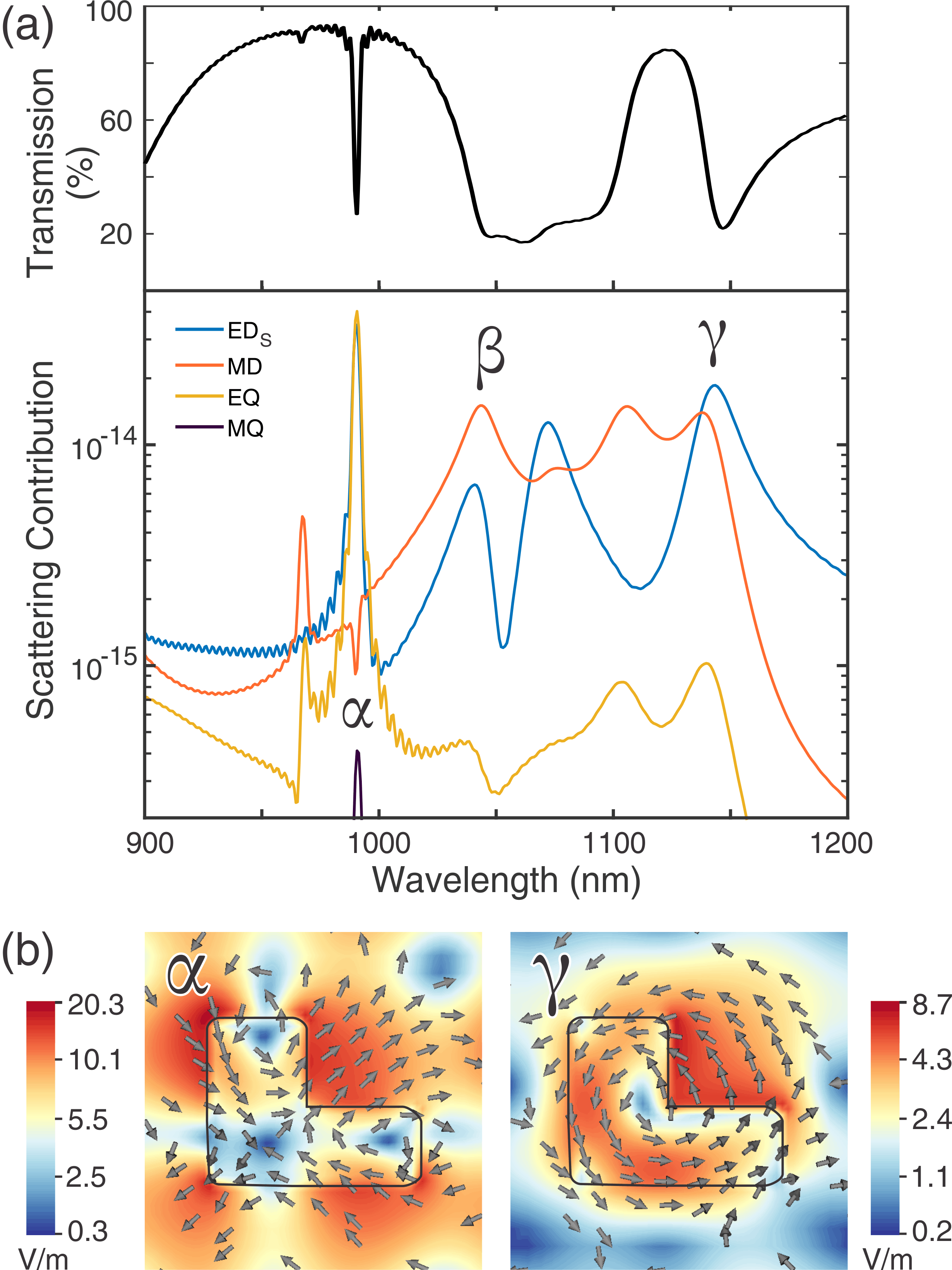}
	\caption{(a) Top: The transmission spectrum of the metasurface with an anisotropic liquid crystal medium and the incident light polarized along the $n_e$ axis.  Bottom: Decomposition into spherical multipoles. (b) The electric field profiles at the frequency of the $\alpha$ and $\gamma$ resonances, respectively.}
    \label{fig:multipoles-lc}
\end{figure}

Due to the LC infiltration, the resonance features of the metasurface shift spectrally and their multipolar origin is affected. Fig.~\ref{fig:multipoles-lc} shows the decomposition into multipoles of a simulation of the metasurface after LC infiltration. The LC was modeled as an anisotropic media which fully surrounds the resonators, and the incident field is polarized along the $n_e$ axis (x-axis in Fig.~\ref{fig:resonator-design}). As a result of the LC infiltration the $\alpha$ resonance is red-shifted from 880\,nm to $\sim990$\,nm and now shows a stronger relative contribution from the electric quadrupole component while the magnetic dipole is suppressed. The important feature to note is that the resonance width is similar to the case of an isotropic medium. The $\beta$ and $\gamma$ resonances however, both show some spectral broadening, especially in the electric dipole contribution to $\gamma$.

\begin{figure}
   \includegraphics[width=\imwidth]{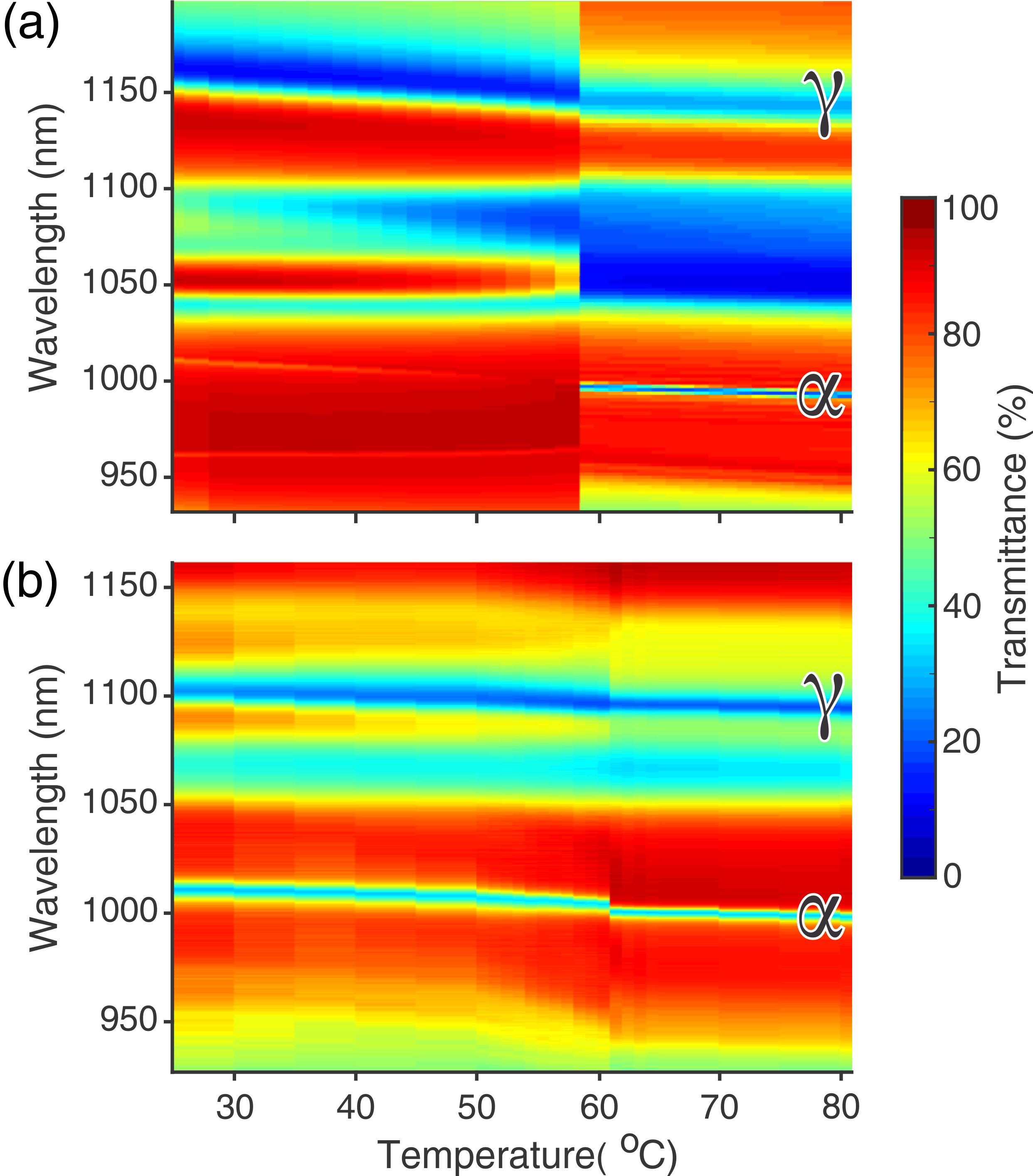}
	\caption{(a) Simulation of the temperature dependence of the transmission spectra of the metasurface. $\alpha$ and $\gamma$ mark the corresponding resonances. (b) Experimentally measured transmittance spectra as a function of temperature. The discontinuity in the plot is due to the transition of the liquid crystals from the nematic to isotropic state.} 
    \label{fig:surface}
\end{figure}

\begin{figure}
   \includegraphics[width=\imwidth]{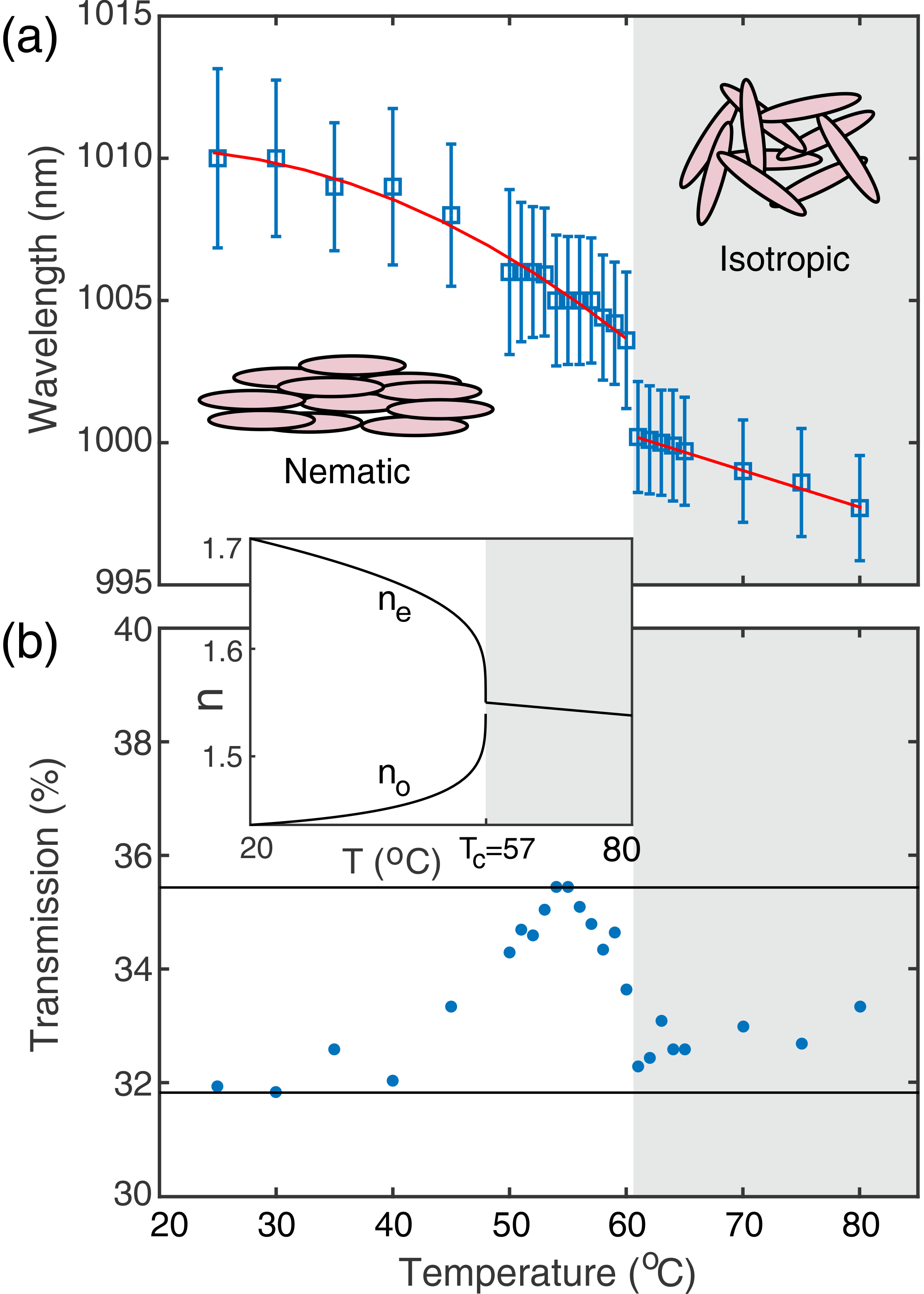}
	\caption{(a) The shift of the $\alpha$ resonance with temperature, where the vertical bars mark the FWHM. The red line is the best fit with a second order polynomial in the nematic state and a linear fit in the isotropic state. (b) The minimum transmission of the resonance as a function of temperature. The inset shows the temperature dependence of the refractive index of E7 liquid crystals at $\lambda=1550$nm~\cite{Li2005b}.}
    \label{fig:fwhm}
\end{figure}

Next, we studied the tunability of the resonances with the change of the LC refractive index, including the change from the nematic to isotropic phase, due to heating. To obtain simulation results that are a function of temperature and to compare our numerical results with the experimental ones we used the known temperature dependence of $n_o$ and $n_e$ of E7 LC~\cite{Li2005b}, where $n_e$ is parallel to to the director of the LCs. Our simulations predict that the resonances will be blue shifted as the medium's temperature increases, as shown in Fig.~\ref{fig:surface}(a). The $\gamma$ resonance is clearly visible in simulation, but the $\alpha$ resonance is predicted to be much finer.

Fig.~\ref{fig:surface}(b) shows the experimental results for the temperature dependence of the transmittance spectra. The heating of the cell was stabilized using a PID control loop on a temperature controller. Because both the $n_o$ and $n_e$ refractive indices of the LCs depend on temperature, we observe a spectral tuning of the resonances due to the heating of the LC. The experimental results show that the $\alpha$ resonance has a constant contrast (maximum to minimum transmission) of approximately $55$ percentage points, which does not dramatically change with temperature. In contrast, the $\gamma$ resonance shows a relatively different contrast at different temperatures, making it less suitable for use in narrow band filtering. The measured  quality factor of the $\alpha$ resonance is $\flatfrac{\lambda_0}{\Delta\lambda}=270\pm30$ in the isotropic LC state, where $\lambda_0$ is the resonance wavelength and $\Delta\lambda$ is the full width at half maximum (FWHM). The figure of merit for the tuning of this resonance is $3.3\pm0.6$. The discontinuity in the graph is the phase transition of the LC from the nematic to isotropic state. The difference between the measured critical temperature of $\sim 61^\circ$C and the listed value of $57^\circ$C for E7 LC is due to temperature gradients across the sample. 

The tuning of the $\alpha$ resonance is made more explicit in Fig.~\ref{fig:fwhm}(a), which shows the resonance wavelength and FWHM as a function of temperature. As can be seen, the shift in the resonance matches the form of the temperature dependence of the refractive index of E7 (Fig.~\ref{fig:fwhm} inset) and thus the shift is due to changes in refractive index of the LCs.  As can be seen in Fig.~\ref{fig:fwhm}(b) the minimum transmission of the resonance changes by less than 4\% across the temperature range measured, thus showing that the resonance remains a good candidate for a band-pass filter across the entire range studied.

In conclusion, we have demonstrated the liquid crystal tuning of high-Q resonant metasurfaces composed of silicon nano-resonators. By introducing asymmetry into the resonator design, we are able to obtain high quality factor transmission features. Specifically, a narrow band transmission dip of only $3.7\pm0.3$~nm width has been shown to be tunable with a figure of merit of $3.3\pm0.6$. All-dielectric metasurfaces therefore offer a feasible path to the construction of tunable blocking filters which would be rugged, lightweight and compact and suitable for various applications, including Raman spectroscopy and hyper spectral imaging.

\textit{Acknowledgments}: 
We would like to thank Yuri Kivshar for the assistance he provided. We acknowledge the financial support by the Australian Research Council and the use of the Australian National Fabrication Facility (ANFF), the ACT Node.
This work was performed, in part, at the Center for Integrated Nanotechnologies, an Office of Science User Facility operated for the U.S. Department of Energy (DOE) Office of Science. Sandia National Laboratories is a multimission laboratory managed and operated by National Technology and Engineering Solutions of Sandia, LLC., a wholly owned subsidiary of Honeywell International, Inc., for the U.S. Department of Energy’s National Nuclear Security Administration under contract DE-NA-0003525. 
\bibliography{physics.bib}

\end{document}